\documentclass[iop]{emulateapj}
\usepackage{graphicx,epsfig,amssymb,amsmath,color,lineno}

\def\eg{{e.g.,~}}
\def\ie{{i.e.,~}}

\begin{document}
\label{firstpage}

\title{MEASUREMENT OF THE EXPANSION RATE OF THE UNIVERSE FROM $\gamma$-RAY ATTENUATION}

\author{{\sc Alberto Dom\'inguez}\altaffilmark{1} and {\sc Francisco Prada}\altaffilmark{2,3,4}}

\slugcomment{Draft; \today}

\shorttitle{The expansion rate of the universe from $\gamma$-ray attenuation}
\shortauthors{DOM\'INGUEZ \& PRADA}

\altaffiltext{1}{Department of Physics \& Astronomy, University of California, Riverside, CA 92521, USA; albertod@ucr.edu}
\altaffiltext{2}{Campus of International Excellence UAM+CSIC, Cantoblanco, E-28049 Madrid, Spain}
\altaffiltext{3}{Instituto de F\'{\i}sica Te\'orica, (UAM/CSIC), Universidad Aut\'onoma de Madrid, Cantoblanco, E-28049 Madrid, Spain}
\altaffiltext{4}{Instituto de Astrof\'{\i}sica de Andaluc\'{\i}a (CSIC), Glorieta de la Astronom\'{\i}a, E-18080 Granada, Spain}



\begin{abstract}
A measurement of the expansion rate of the universe (that is the Hubble constant, $H_{0}$) is derived here using the $\gamma$-ray attenuation observed in the spectra of $\gamma$-ray sources produced by the interaction of extragalactic $\gamma$-ray photons with the photons of the extragalactic background light (EBL). The Hubble constant that is determined with our technique, for a $\Lambda$CDM cosmology, is $H_{0}=71.8_{-5.6}^{+4.6}({\rm stat})_{-13.8}^{+7.2}({\rm syst})$~km~s$^{-1}$~Mpc$^{-1}$. This value is compatible with present-day measurements using well established methods such as local distance ladders and cosmological probes. The recent detection of the cosmic $\gamma$-ray horizon (CGRH) from multiwavelength observation of blazars, together with the advances in the knowledge of the EBL, allow us to measure the expansion rate of the universe. This estimate of the Hubble constant shows that $\gamma$-ray astronomy has reached a mature enough state to provide cosmological measurements, which may become more competitive in the future with the construction of the Cherenkov Telescope Array. We find that the maximum dependence of the CGRH on the Hubble constant is approximately between redshifts 0.04 and 0.1, thus this is a smoking gun for planning future observational efforts. Other cosmological parameters, such as the total dark matter density $\Omega_{m}$ and the dark energy equation of state $w$, are explored as well. 
\end{abstract}

\keywords{BL Lacertae objects: general --- cosmic background radiation --- cosmological parameters --- cosmology: observations, diffuse radiation }

\section{Introduction}
\label{sec:intro}
The universe is not transparent to very high energy (VHE) photons (30~GeV--300~TeV) traveling through cosmological distances. A $\gamma$-ray photon and an extragalactic background light (EBL) photon in the intergalactic medium, mainly produced by star formation throughout the cosmic history of the universe, may annihilate and produce an electron-positron pair (\citealt{nikishov62,gould66,stecker92}). This process generates an attenuation in the spectra of $\gamma$-ray sources above a critical $\gamma$-ray energy, which has been observed with the current generation of $\gamma$-ray telescopes (\citealt{ackermann12,abramowski13,dominguez13}, hereafter D13).

Recently, D13 presented the first detection of the cosmic $\gamma$-ray horizon (CGRH). The CGRH is by definition the energy as a function of redshift at which the intrinsic flux emitted by the source, that we would observe without EBL attenuation, decreases $1/e$ (or also by approximately 65\%) due to interactions with the EBL. The detection of the CGRH has been possible thanks to the recent data collected by the \emph{Fermi} satellite and multiwavelength observations for a dozen of blazars that include detections by Imaging Atmospheric Cherenkov Telescopes (IACTs). This observational measurement of the CGRH allows the comparison with predictions from robust EBL models, which leads for the first time to measuring cosmological parameters in $\gamma$-ray astronomy.



In this Letter, we measure the local expansion rate of the universe $H_{0}$ from $\gamma$-ray attenuation leading to a good agreement with well established methods such as local distance ladders as well as cosmological measurements. Our estimate of the Hubble constant is completely independent of those methods mentioned above, and it may become competitive when more data is available, specially in the era of the upcoming Cherenkov Telescope Array (CTA). As discussed by \citet{suyu12}, multiple paths to independent determinations of the Hubble constant are needed in order to access and control systematic uncertainties. Accurate estimates of $H_{0}$ provide critical independent constraints on dark energy, spatial curvature, neutrino physics, and general relativity (\citealt{freedman10,suyu12,weinberg12}).


This Letter is organized as follows: \S\ref{sec:dependence} describes how to empirically derive the CGRH assuming different values of the cosmological parameters. In \S\ref{sec:cosmo}, we constrain the value of the Hubble constant that is favored by the observed CGRH. Finally, \S\ref{sec:summary} presents a discussion and summary of our results.

\section{The $\gamma$-ray attenuation dependence on the cosmological parameters} \label{sec:dependence}
The $\gamma$-ray optical depth $\tau$ produced by the pair production interaction between a $\gamma$-ray photon and an EBL photon is analytically given by

\begin{equation}
\label{attenu}
\tau(E,z)=\int_{0}^{z} \Big(\frac{dl}{dz'}\Big) dz' \int_{0}^{2}d\mu \frac{\mu}{2}\int_{\varepsilon_{th}}^{\infty} d\varepsilon{'}\ \sigma_{\gamma\gamma}(\beta{'})n(\varepsilon{'},z').
\end{equation}

The lower limit of the energy integral $\varepsilon_{th}$ is the energy threshold of the pair production interaction that is explicitly given by

\begin{equation}
\label{threshold}
\varepsilon_{th}\equiv \frac{2m_{e}^2c^{4}}{E\mu},
\end{equation}
\noindent where $E'$ is the energy of the $\gamma$ photon (in the rest-frame at redshift $z'$), $\varepsilon{'}$ is the energy of the EBL photon (in the rest-frame at redshift $z'$), and $\mu=(1-\cos \theta)$, with $\theta$ the angle of the interaction. The constant $m_{e}$ is the electron mass and $c$ the vacuum speed of light.

The factor $n(\varepsilon{'},z')$ in Equation~(\ref{attenu}) is the proper number density per unit energy of EBL photons and the parameter $\sigma_{\gamma\gamma}$ is the photon-photon pair production cross section and $\beta{'}$ is

\begin{equation}
\label{beta}
\beta^{'}=\frac{\varepsilon_{th}}{\varepsilon{'}(1+z')^{2}}.
\end{equation}

The factor $dl/dz'=c|dt/dz'|$ in Equation~(\ref{attenu}) defines how the infinitesimal space element varies with redshift, which according to \citet{peebles93} is given by

\begin{equation} 
\label{peebles}
\Big|\frac{dt}{dz'}\Big|=\frac{1}{H_{0}(1+z')E(z')}
\end{equation}
\noindent with

\begin{equation}
\label{eq:Ez}
E(z') \equiv \sqrt{\Omega_{m}(1+z')^3+\Omega_{\Lambda}} \, \, , 
\end{equation}

\noindent and $H_{0}$, $\Omega_{m}$ and $\Omega_{\Lambda}$ given by the parameters of the flat $\Lambda$CDM cosmology.\\

Given the exponential flux attenuation produced by the EBL, the CGRH may be defined as the energy $E_{0}$ as a function of redshift at which $\tau(E_{0},z)=1$. From Equation~(\ref{attenu}), we see that $\tau$ is dependent on the cosmological parameters by two factors. First, the dependence given by the EBL density evolution $n(\varepsilon,z)$. Second, the dependence with the extragalactic $\gamma$-ray propagation through the universe given by the factor $dl/dz$. Interestingly, both effects contribute quantitatively as well as qualitatively differently to $\tau$. These two factors are discussed thoroughly in the following subsections \S\ref{subsec:n} and \S\ref{subsec:interplay}. The net effect in $\tau$ of varying $H_{0}$ is an interplay between the contribution from these factors, which is also described in \S\ref{subsec:interplay}.

\begin{figure*}
\includegraphics[width=18cm]{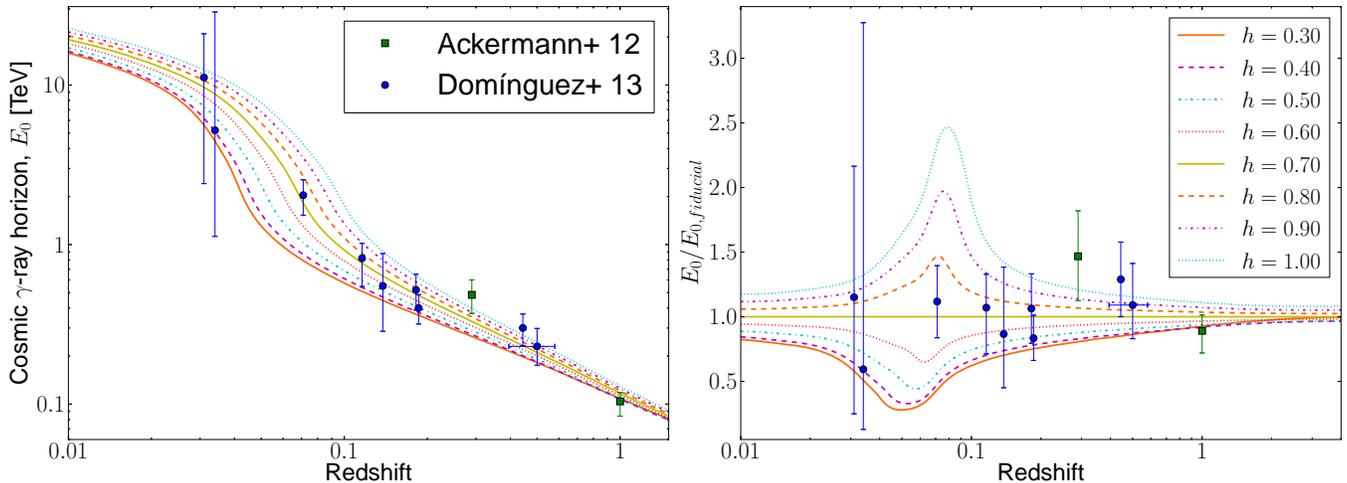}
\caption{Left panel: the CGRH for different values of the Hubble constant, as predicted from the empirical EBL modeling by D11 described in the text, are shown with several line styles and colors (a flat $\Lambda$CDM cosmology with matter density $\Omega_{m}=0.3$ is assumed). The CGRH data are taken from Ackermann et al. (2012, filled green squares) and D13 (filled blue circles). The error bars include the total uncertainty (statistical plus systematic). Right panel: same as left panel but all the $E_{0}$ values have been normalized to the empirical CGRH derived for the fiducial cosmology with $h=0.7$ and $\Omega_{m}=0.3$.}
\label{fig1}
\end{figure*}

\subsection{Extragalactic background light photon evolution}\label{subsec:n}

Here, we adopt the methodology described in Dom\'inguez et al. (2011, hereafter D11) to compute the evolving EBL (which is given by $n$ in Equation~(\ref{attenu})) for different values of $h$ (the dimensionless parameter $h=H_{0}/100$), whereas $\Omega_{m}$ (and hence $\Omega_{\Lambda}\equiv 1-\Omega_{m}$ for the adopted flat $\Lambda$CDM cosmology), have been fixed to $\Omega_{m}=0.3$. This choice is compatible with the latest constraints \ie $\Omega_{m}=0.307\pm 0.010$ (\citealt{ade13}), and with the dark matter density adopted in the spectral energy distribution (SED) analysis of the blazars used by D13 to measure the CGRH. As we will see in \S\ref{sec:cosmo}, the choice of $\Omega_{m}$ is not significantly relevant for our results.


The EBL was empirically derived by D11 from two main ingredients. First, the estimation of galaxy SED-type fractions based upon a multiwavelength catalog of around 6,000 galaxies drawn from the All-wavelength Extended Groth strip International Survey (\citealt{davis07}). We have checked that the SED-type fractions have very little dependence with the cosmological parameters.

Second, the $K$-band galaxy luminosity functions (LFs) from \citet{cirasuolo10}. This observable gives the number of galaxies per unit volume and magnitude in the near-IR, from the local universe up to $z\sim 4$. The galaxy LFs are described by Schechter functions (\citealt{schechter76}), whose parameters depend on cosmology. Analytically, for a given cosmology, the Schechter functions are parameterized by three quantities: $\phi_{0}(z)$, $M_{*}(z)$, and $\alpha$ (the normalization, a characteristic absolute magnitude, and the faint-end slope). It is then possible to compute the Schechter LFs for a new set of  $\Lambda$CDM cosmological parameters ($h$, $\Omega_{m}$, $\Omega_{\Lambda}$) providing the values of $\phi^{'}_{0}(z)$ and $M^{'}_{*}(z)$ obtained adopting the fiducial parameters ($h'$, $\Omega_{m}^{'}$ and $\Omega_{\Lambda}^{'}$). Below, we provide the equations to convert the Schechter LF, written in absolute magnitudes, from a fiducial set of cosmological parameters to another choice, i.e.

\[
\phi(M,h,\Omega_{m},\Omega_{\lambda},z)=0.4\ln(10)\phi_{0} \times 10^{0.4(M_{*}-M)(\alpha+1)}
\]
\begin{equation}
\label{eq:last}
\hspace{0.5cm}\times \exp[-10^{0.4(M_{*}-M)}]  \hspace{1.6cm}\textrm{[Mpc$^{-3}$~Mag$^{-1}$]}
\end{equation}

\begin{equation}
\phi_{0}=\phi_{0}'\Big(\frac{h}{h'}\Big)^{3} \frac{E(z)}{E'(z)} \Big(\frac{F'(z)}{F(z)}\Big)^{2}
\end{equation}

\begin{equation}
M_{*}=M_{*}'+5\log_{10}\Big[\frac{h}{h'}\frac{F'(z)}{F(z)}\Big]
\end{equation}

\noindent where

\begin{equation} \label{eq:F}
F(z) = \int_{0}^{z}\frac{dz'}{E(z')},
\end{equation}

\noindent with $E(z')$ given by Equation~(\ref{eq:Ez}). We note that Equation~(\ref{eq:F}) is proportional to the time derivative of the logarithm of the scale factor, with $z$ redshift and $\Omega_{m}$ and $\Omega_{\Lambda}$ density parameters. For this reason, $H(z) = H_{0} E(z)$ is the Hubble constant as measured by an observer located at $z$ (see \citealt{hogg99}). We recall that quantities noticed with $'$ correspond to that adopting fiducial cosmological parameters. In our equations, we assume that the power-law $\alpha$ parameter of the Schechter LF does not depend on cosmology, which is a reasonable assumption as it can be seen from the Schechter fits to galaxy data in Fried et al. (2001, see Table 2, \eg for $\Omega_{m}=0.3$ and $\Omega_{m}=1$).

Therefore, from Equation~(\ref{eq:last}) and following the methodology described in D11 is possible to calculate the luminosity densities, and thus the EBL evolution for different cosmologies. 

\subsection{Propagation of the $\gamma$-ray photons and cosmological dependence on the optical depth $\tau$} \label{subsec:interplay}


As we mentioned above, the dependence of the CGRH on cosmology comes from two factors. First, from the dependence of the EBL evolution with cosmology ($n$ in Equation~(\ref{attenu}) and described in \S\ref{subsec:n}). Second, the factor that account for the propagation of the $\gamma$-ray photon through space ($dl/dz$ in Equation~(\ref{attenu})). Both are quantitatively and qualitatively different.

First, the $n$ factor in $\tau$ depends proportionally on the Hubble constant. Therefore, $\tau$ is higher at a given $\gamma$-ray energy and redshift for larger values of $h$. The dependence is not linear with $h$. For example, the local EBL spectral intensity estimated for $h=0.2$ is approximately three times larger than the intensity derived for $h=0.1$. However, the local EBL intensity estimated for $h=1$ is only two times larger than that derived for $h=0.5$.

The other factor $dl/dz$ goes in the opposite direction since this factor is inversely proportional to the Hubble constant. This means that $\tau$ is smaller at a given $\gamma$-ray energy and redshift for larger values of $h$. This is understandable since $h$ is a measure of how fast the universe is expanding. Therefore, when the universe expands at a slower rate, VHE photons travel a shorter distance and thus the probability that those $\gamma$ photons interact with the EBL is indeed lower. This is a linear factor and dominates for $h\gtrsim 0.3$, producing the net effect that the universe is more transparent (lower $\tau$) for larger values of the Hubble constant. For $h\lesssim 0.3$, the $n$ factor takes over and produces an inversion of this trend. For instance, the universe would have a maximum opacity for $h\sim 0.3$.

\section{Measuring the Hubble constant from $\gamma$-ray attenuation} \label{sec:cosmo}

\subsection{Theoretical and observational background}

The potential of measuring the Hubble constant from $\gamma$-ray attenuation was already pointed out two decades ago by \citet{salamon94} and \citet{mann96}, when the $\gamma$-ray experiments at that time could only study a few sources on the entire sky. In the last decade, \citet{blanch05a,blanch05b,blanch05c} studied, in a series of papers, the potential of using the CGRH to constrain cosmology. These investigations were motivated by the starting operation of the new IACTs such as H.E.S.S., MAGIC, and VERITAS (\citealt{hinton04,lorenz04,weekes02}, respectively). Blanch \& Mart\'inez used simulated VHE spectra of blazars, at different redshifts, to estimate how relevant cosmological parameters could be constrained. Their analysis was based on the fact that the CGRH depends on the propagation of the VHE photons through cosmological distances, which is dependent on cosmology. Yet, they neglected the contribution on the cosmological dependence encoded in the evolution of the EBL spectral intensity with redshift. These two effects are consistently considered in our analysis. \citet{barrau08} also understood the potential of $\gamma$-ray attenuation to constrain cosmological parameters. They derive a lower limit of the Hubble constant, $H_{0}>74$~km~s$^{-1}$~Mpc$^{-1}$ at a 68\% confidence level, from the observation of $\gamma$-ray photons coming from a flare of the blazar Mkn~501, which was detected by HEGRA (\citealt{aharonian99}).

Independently, the knowledge of the EBL has largely improved in the last few years (see for a review \citealt{primack11}, \citealt{dominguez12}, and \citealt{dwek13}). Recently, direct measurements in optical wavelengths of the EBL in the local universe (\citealt{matsuoka11,mattila12}) have confirmed previous indications (\eg \citealt{aharonian06}) of an EBL intensity level close to the estimations from deep galaxy counts (\eg \citealt{madau00,keenan10}). Furthermore, realistic EBL models based on large multiwavelength galaxy data sets such as the one found in D11 and a better theoretical understanding of galaxy evolution (\eg \citealt{somerville12,gilmore12}) have allowed both, the understanding of the EBL at wavelengths where the detection is not possible yet and the convergence of different methodologies. 

\begin{figure}
\includegraphics[width=\columnwidth]{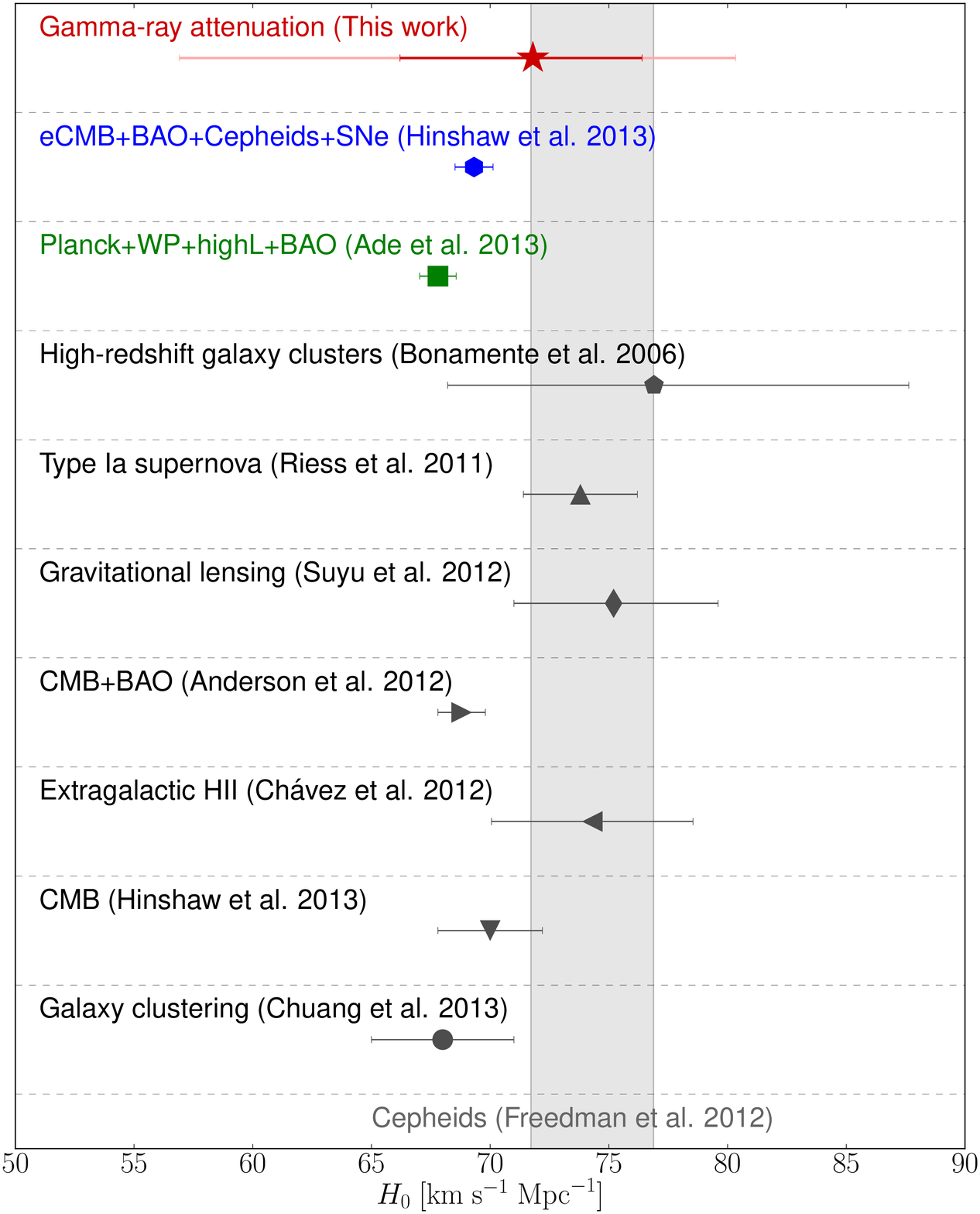}
\caption{The Hubble constant $H_{0}$ derived from different methodologies. The measurement presented in this work is shown with a red star. For this measurement, the statistical uncertainties are shown with darker red whereas the total uncertainties (statistical plus systematic, added in quadrature) are shown with lighter red. The combined value presented by \citet{hinshaw12} is shown with a blue hexagon, which include CMB data from WMAP9 plus the ground-based SPT and ACT (extended CMB or eCMB), BAO, and Cepheids plus SNe measurements. The CMB+BAO measurement by \citet{anderson12} includes CMB data from WMAP7 and BAO from the SDSS-II luminous red galaxy sample  plus data from the Baryon Oscillation Spectroscopic Survey (BOSS). The results from the Planck Space Telescope combined with WMAP polarization low-multipole likelihood (WP) plus high-resolution CMB data (highL and BAO, \citealt{ade13}) are shown with a green square. As a reference, a shaded region is showing the $H_{0}$ value from the Cepheids distance ladder.}
\label{fig2}
\end{figure}


\subsection{Methodology}

We base our estimation of the Hubble constant on the hypothesis that the evolving EBL is sufficiently well described by the model presented in D11. This choice is supported, as mentioned above, by independent observational data sets and the convergence of EBL models using different methodologies. The uncertainties in the EBL model, which are estimated by D11, are also taken into account in our cosmological analysis. We stress that the CGRH derived in the relevant redshift range from other EBL models such as those from \citet{franceschini08}, \citet{finke10}, and \citet{gilmore12} are within the uncertainties of the D11 model.

The CGRH derived following the D11 EBL methodology but adopting different values of the Hubble constant, for a flat $\Lambda$CDM universe with a fixed matter density $\Omega_M=0.3$, is shown in Figure~\ref{fig1} (left panel). We set the uniform prior that $0.3\leq h\leq 1$ in agreement with other observational constrains. This choice is made to avoid the inversion of the trend for $h\lesssim 0.3$ described in \S\ref{subsec:interplay}, which makes that the overall likelihood distribution has two maxima: a global maximum at $h\sim 0.1$ and the value of the Hubble constant that we report. As discussed in \S\ref{sec:dependence}, we notice that, in the explored $H_{0}$ range, the universe is more transparent to VHE photons for lower values of the Hubble constant. Figure~\ref{fig1} also shows the CGRH data presented in \citet{ackermann12} and D13. \citet{ackermann12} stack hundreds of spectra from blazars detected by the \emph{Fermi} satellite in order to search for an EBL attenuation feature. They do not provide directly any results in terms of the CGRH, but this can be estimated from their Figure~2 taking the average redshift of the bin and the energy value where $\exp(-\tau)=1/e$ (M. Ajello, private communication). We note that this energy is not currently probed by \emph{Fermi} for their lowest redshift bin ($z<0.2$). The error bars shown by D13 are the total statistical plus systematic uncertainties, which are added in quadrature.


Our analysis is based on applying a maximum likelihood technique in order to find which CGRH models (and therefore which Hubble constant) are favored by the CGRH data. In this analysis, the systematic uncertainties in the determination of the Hubble constant are considered as well. These are measured by applying our maximum likelihood methodology to the cases that bracket the evolving EBL uncertainties given in the D11 model. 

\begin{figure}
\includegraphics[width=\columnwidth]{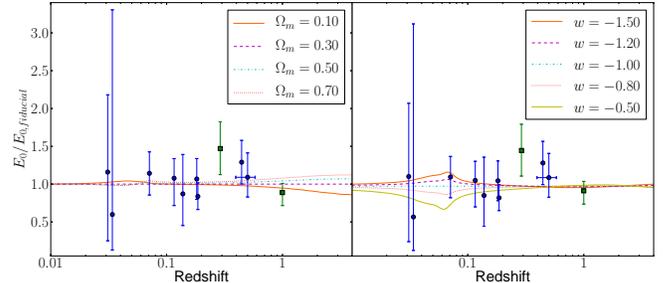}
\caption{The CGRH derived empirically up to $z=4$ normalized by the CGRH for the fiducial case of a flat universe with $h=0.7$, $\Omega_{m}=0.3$, and $w=-1$. Left panel: several values of the matter density $\Omega_{m}$. Right panel: several values of the dark energy equation of state $w$. The data from \citet{ackermann12} (green squares) and D13 (blue circles) are shown.}
\label{fig3}
\end{figure}

Figure~\ref{fig1} (right panel) shows the predicted CGRH estimated adopting different values of the Hubble constant but normalized at the values obtained for the CGRH model with fiducial cosmological parameters $h=0.7$, $\Omega_{m}=0.3$ and $\Omega_{m}=0.7$. This figure is intended to highlight the largest dependence with redshift. Hence, we can see that the highest sensitivity to the Hubble constant is approximately in the range from redshift $0.04$ to $0.1$. This is a smoking gun to plan upcoming IACT observations and analysis of VHE sources, which will yield competitive constrains on $H_{0}$. The reason for this optimal redshift is that this is the window where $\tau=1$ at energies where the optical depth flattens with energy (see Figure~17 in D11). For lower redshifts, this occurs at $\tau<1$ and for higher redshifts at $\tau>1$. The flattening in $\tau$ produces that small changes in $h$ imply significant variations in $E_{0}$.

In the present work, the best-fit CGRH model to the actual data yield a value of the Hubble constant of $H_{0}=71.8_{-5.6}^{+4.6}({\rm stat})_{-13.8}^{+7.2}({\rm syst})$~km~s$^{-1}$~Mpc$^{-1}$. In applying this procedure we have to assume that the uncertainties of the CGRH data (that include systematic uncertainties in the \emph{Fermi}-Large Area Telescope energy scale, see D13) are distributed as a Gaussian, which is not necessarily true. In the likelihood fit, the width $\sigma$ of the Gaussian, this is, $\propto \exp(-(x-\mu)/(2\sigma^{2}))$, is assumed as the mean value between the lower and upper uncertainty of the data being fitted. As explained above, the statistical errors are derived from the maximum likelihood fit and the systematic uncertainties are then accounted for the EBL modeling. We notice that the EBL model uncertainties are asymmetric (see \S6.1 in D11) and therefore also the systematic uncertainties in the Hubble constant estimation.

The value of $H_0$ obtained here, based on $\gamma$-ray attenuation, with a total accuracy of about $18\%$, is in good agreement with that of other present-day methods, as shown in Figure~\ref{fig2}. Our estimate is compared with the Hubble constant and its uncertainties obtained using the Cepheids (\citealt{freedman12}), type Ia supernovae (SNe, \citealt{riess11}), and extragalactic HII regions (\citealt{chavez12}) distance ladders, as well as that provided by cosmological probes such as the latest results from the cosmic microwave background (CMB) data, \ie Atacama Cosmology Telescope (ACT, \citealt{fowler10}), South Pole Telescope (SPT, \citealt{keisler11}), Wilkinson Microwave Anisotropy Probe (WMAP9, \citealt{hinshaw12}) and Planck Space Telescope (\citealt{ade13}), BOSS galaxy clustering (\citealt{chuang13}), baryonic acoustic oscillations (BAO, \citealt{anderson12}), time-delay strong gravitational lensing (\citealt{suyu13}) and Sunyaev-Zel'dovich effect plus X-ray measurements of high-redshift galaxy clusters (\citealt{bonamente06}). We also show the results obtained from the combined measurement using different techniques, which are taken from \citet{hinshaw12} and \citet{ade13}, see Figure~\ref{fig2}.


From our methodology, it is possible to test the dependence of the CGRH with other cosmological parameters such as the matter density $\Omega_{m}$. The procedure is the same as that adopted in the case of the Hubble constant. Now, the Hubble constant is being fixed and $\Omega_{m}$ is varied. The results can be seen in the left panel of Figure~\ref{fig3}. The same procedure can be taken in order to determine the dependence of the CGRH on the dark energy equation of state $w$ by substituting $\Omega_{\Lambda}$ in Equation~(\ref{eq:Ez}), by $\Omega_{\Lambda}(1+z)^{3(1+w)}$. These results are shown in the right panel of Figure~\ref{fig3}. In both cases $\Omega_{m}$ and $w$, we find that the CGRH does not significantly depend on these parameters and their constrains from $\gamma$-ray attenuation are hardly feasible. 

\section{Discussion and conclusions} \label{sec:summary}
In this Letter we have demonstrated the degree of maturity accomplished in $\gamma$-ray astronomy on measuring, for the first time, the Hubble constant; which is in good agreement with present-day distance ladder methods and cosmological probes. This has been possible thanks to the new generation of IACT telescopes and the \emph{Fermi} satellite, combined with multiwavelength observations of a sample of well-studied blazars up to $z=1$ plus the advances in the EBL knowledge.



The most likely value of $H_{0}$ that we found is $H_{0}=71.8_{-5.6}^{+4.6}({\rm stat})_{-13.8}^{+7.2}({\rm syst})$~km~s$^{-1}$~Mpc$^{-1}$. The total uncertainties in the estimate of $H_{0}$ are dominated by the uncertainties propagated from the systematics in the D11 EBL model. These uncertainties will be reduced mainly with a better understanding of the galaxy SEDs from the ultraviolet to far infrared wavelengths. The accuracy of our methodology is significantly asymmetric due to the propagation of the asymmetry in the EBL model as described in D11. The mean total uncertainty in the value of $H_{0}$ is estimated as $\sim 18\%$, whereas if we consider only the statistical uncertainties, the accuracy improves to a mean of $\sim 8\%$. This accuracy is not as high as other techniques such as Cepheids, SNe, CMB, or BAO, which are of the order of $\sim 3\%$. Yet, this is the first time that $\gamma$-ray attenuation data are used to measure the expansion rate of the universe. The prospects of increasing the accuracy of $H_{0}$ with this technique are promising as our understanding of the EBL improves and new CGRH data will become available. Specially, we emphasize as a result of our work that observations in the redshift range $z\sim 0.04$--$0.1$ will improve substantially the $H_{0}$ estimate. These data will be available mainly thanks to the increase of simultaneous multiwavelength campaigns and high energy data from the \emph{Fermi} satellite, current IACTs, and specially from the upcoming CTA experiment. 

\section*{Acknowledgments}
We thank Marco Ajello, Juan Cortina, Justin Finke, Wendy Freedman, Barry Madore, Joel Primack, Brian Siana and the anonymous referee for helpful comments. We also thank Gillian Wilson and Nathaniel Stickley for providing computational resources. We acknowledge the support of the Spanish MICINN's Consolider-Ingenio 2010 Programme under grant MultiDark CSD2009-00064.

\bibliographystyle{plain}

\begin{thebibliography}{}










\bibitem[\protect\citeauthoryear{Abramowski et al.}{2013}]{abramowski13} Abramowski A., et al., 2013, A\&A, 550, A4







\bibitem[\protect\citeauthoryear{Ackermann et al.}{2012}]{ackermann12} Ackermann M., et al., 2012, Science, 338, 1190

\bibitem[\protect\citeauthoryear{Ade et al.}{2013}]{ade13} Ade P.~A.~R., et al., 2013, arXiv, arXiv:1303.5076 

\bibitem[\protect\citeauthoryear{Aharonian et al.}{1999}]{aharonian99} Aharonian F., et al., 1999, A\&A, 349, 11 




\bibitem[\protect\citeauthoryear{Aharonian et al.}{2006}]{aharonian06} Aharonian F., et al., 2006, Nature, 440, 1018













\bibitem[\protect\citeauthoryear{Anderson et al.}{2012}]{anderson12} Anderson L., et al., 2012, MNRAS, 427, 3435 




\bibitem[\protect\citeauthoryear{Barrau, Gorecki \& Grain}{2008}]{barrau08} Barrau A., Gorecki A., Grain J., 2008, MNRAS, 389, 919 







\bibitem[\protect\citeauthoryear{Blanch \& Mart\'inez}{2005a}]{blanch05a} Blanch O., Martinez M., 2005a, APh, 23, 588

\bibitem[\protect\citeauthoryear{Blanch \& Mart\'inez}{2005b}]{blanch05b} Blanch O., Martinez M., 2005b, APh, 23, 598

\bibitem[\protect\citeauthoryear{Blanch \& Mart\'inez}{2005c}]{blanch05c} Blanch O., Martinez M., 2005c, APh, 23, 608






\bibitem[\protect\citeauthoryear{Bonamente et al.}{2006}]{bonamente06} Bonamente M., Joy M.~K., LaRoque S.~J., Carlstrom J.~E., Reese E.~D., Dawson K.~S., 2006, ApJ, 647, 25 












\bibitem[\protect\citeauthoryear{Ch{\'a}vez et al.}{2012}]{chavez12} Ch{\'a}vez R., Terlevich E., Terlevich R., Plionis M., Bresolin F., Basilakos S., Melnick J., 2012, MNRAS, 425, L56 

\bibitem[\protect\citeauthoryear{Chuang et al.}{2013}]{chuang13} Chuang A., et al., 2013, in preparation.

\bibitem[\protect\citeauthoryear{Cirasuolo et al.}{2010}]{cirasuolo10} Cirasuolo M., McLure R.~J., Dunlop J.~S., Almaini O., Foucaud S., Simpson C., 2010, MNRAS, 401, 1166











\bibitem[\protect\citeauthoryear{Davis et al.}{2007}]{davis07} Davis M., et al., 2007, ApJ, 660, L1





\bibitem[\protect\citeauthoryear{Dom\'inguez et al.}{2011a}]{dominguez11a} Dom\'inguez A., et al., 2011, MNRAS, 410, 2556 (D11)


\bibitem[\protect\citeauthoryear{Dom{\'{\i}}nguez}{2012}]{dominguez12} Dom{\'{\i}}nguez A., 2012, IAUS, 284, 442


\bibitem[\protect\citeauthoryear{Dom\'inguez et al.}{2013}]{dominguez13} Dom\'inguez A., et al., 2013, ApJ, 770, 77 (D13)




\bibitem[\protect\citeauthoryear{Dwek \& Krennrich}{2013}]{dwek13} Dwek E., Krennrich F., 2013, APh, 43, 112 











\bibitem[\protect\citeauthoryear{Finke, Razzaque \& Dermer}{2010}]{finke10} Finke J.~D., Razzaque S., Dermer C.~D., 2010, ApJ, 712, 238

\bibitem[\protect\citeauthoryear{Fowler et al.}{2010}]{fowler10} Fowler J.~W., et al., 2010, ApJ, 722, 1148 

\bibitem[\protect\citeauthoryear{Franceschini, Rodighiero \& Vaccari}{2008}]{franceschini08} Franceschini A., Rodighiero G., Vaccari M., 2008, A\&A, 487, 837

\bibitem[\protect\citeauthoryear{Freedman \& Madore}{2010}]{freedman10} Freedman W.~L., Madore B.~F., 2010, ARA\&A, 48, 673 

\bibitem[\protect\citeauthoryear{Freedman et al.}{2012}]{freedman12} Freedman W.~L., Madore B.~F., Scowcroft V., Burns C., Monson A., Persson S.~E., Seibert M., Rigby J., 2012, ApJ, 758, 24 

\bibitem[\protect\citeauthoryear{Fried et al.}{2001}]{fried01} Fried J.~W., et al., 2001, A\&A, 367, 788








\bibitem[\protect\citeauthoryear{Gilmore et al.}{2012}]{gilmore12} Gilmore R.~C., Somerville R.~S., Primack J.~R., Dom{\'{\i}}nguez A., 2012, MNRAS, 422, 3189 



\bibitem[\protect\citeauthoryear{Gould \& Schr{\'e}der}{1966}]{gould66} Gould R.~J., Schr{\'e}der G., 1966, PhRvL, 16, 252





\bibitem[\protect\citeauthoryear{Hinshaw et al.}{2012}]{hinshaw12} Hinshaw G., et al., 2012, arXiv, arXiv:1212.5226 

\bibitem[\protect\citeauthoryear{Hinton}{2004}]{hinton04} Hinton J.~A., 2004, NewAR, 48, 331







\bibitem[\protect\citeauthoryear{Hogg}{1999}]{hogg99} Hogg D.~W., 1999, astro, arXiv:astro-ph/9905116 






\bibitem[\protect\citeauthoryear{Keenan et al.}{2010}]{keenan10} Keenan R.~C., Barger A.~J., Cowie L.~L., Wang W.-H., 2010, ApJ, 723, 40 

\bibitem[\protect\citeauthoryear{Keisler et al.}{2011}]{keisler11} Keisler R., et al., 2011, ApJ, 743, 28













\bibitem[\protect\citeauthoryear{Lorenz}{2004}]{lorenz04} Lorenz E., 2004, NewAR, 48, 339


\bibitem[\protect\citeauthoryear{Madau \& Pozzetti}{2000}]{madau00} Madau P., Pozzetti L., 2000, MNRAS, 312, L9

\bibitem[\protect\citeauthoryear{Mannheim}{1996}]{mann96} Mannheim K., 1996, RvMA, 9, 17






\bibitem[\protect\citeauthoryear{Matsuoka et al.}{2011}]{matsuoka11} Matsuoka Y., Ienaka N., Kawara K., Oyabu S., 2011, ApJ, 736, 119 



\bibitem[\protect\citeauthoryear{Mattila et al.}{2012}]{mattila12} Mattila K., Lehtinen K., V{\"a}is{\"a}nen P., von Appen-Schnur G., Leinert C., 2012, IAUS, 284, 429 












\bibitem[\protect\citeauthoryear{Nikishov}{1962}]{nikishov62} Nikishov A.~I., 1962, Sov. Phys. JETP, 14, 393





\bibitem[\protect\citeauthoryear{Peebles}{1993}]{peebles93} Peebles P.~J.~E., 1993, Principles of Physical Cosmology, Princeton University Press








\bibitem[\protect\citeauthoryear{Primack et al.}{2011}]{primack11} Primack J.~R., Dom{\'{\i}}nguez A., Gilmore R.~C., Somerville R.~S., 2011, AIPC, 1381, 72 




\bibitem[\protect\citeauthoryear{Riess et al.}{2011}]{riess11} Riess A.~G., et al., 2011, ApJ, 730, 119


\bibitem[\protect\citeauthoryear{Salamon, Stecker \& de Jager}{1994}]{salamon94} Salamon M.~H., Stecker F.~W., de Jager O.~C., 1994, ApJ, 423, L1 







\bibitem[\protect\citeauthoryear{Schechter}{1976}]{schechter76} Schechter P., 1976, ApJ, 203, 297











\bibitem[\protect\citeauthoryear{Somerville et al.}{2012}]{somerville12} Somerville R.~S., Gilmore R.~C., Primack J.~R., Dom{\'{\i}}nguez A., 2012, MNRAS, 423, 1992 

\bibitem[\protect\citeauthoryear{Stecker, de Jager \& Salamon}{1992}]{stecker92} Stecker F.~W., de Jager O.~C., Salamon M.~H., 1992, ApJ, 390, L49





\bibitem[\protect\citeauthoryear{Suyu et al.}{2012}]{suyu12} Suyu S.~H., et al., 2012, arXiv, arXiv:1202.4459

\bibitem[\protect\citeauthoryear{Suyu et al.}{2013}]{suyu13} Suyu S.~H., et al., 2013, ApJ, 766, 70 













\bibitem[\protect\citeauthoryear{Weekes et al.}{2002}]{weekes02} Weekes T.~C., et al., 2002, APh, 17, 221

\bibitem[\protect\citeauthoryear{Weinberg et al.}{2012}]{weinberg12} Weinberg D.~H., Mortonson M.~J., Eisenstein D.~J., Hirata C., Riess A.~G., Rozo E., 2012, arXiv, arXiv:1201.2434 






\end{thebibliography}

\label{lastpage}
\end{document}